\documentstyle [12pt,A4,psfig] {article}
\def\cen{\centerline}
\begin{document}

\cen{\Large{\bf The Observation of Up-going Charged Particles Produced }}\vskip 
0.3 truecm
\cen{\Large{\bf by High Energy Muons in Underground Detectors}}\vskip 1.3 truecm

\bigskip\begin{center}
{\bf The MACRO Collaboration}\\
\nobreak\bigskip\nobreak
\pretolerance=10000
M.~Ambrosio$^{12}$, 
R.~Antolini$^{7}$, 
C.~Aramo$^{7,p}$,
G.~Auriemma$^{14,a}$, 
A.~Baldini$^{13}$, 
G.~C.~Barbarino$^{12}$, 
B.~C.~Barish$^{4}$, 
G.~Battistoni$^{6,b}$, 
R.~Bellotti$^{1}$, 
C.~Bemporad$^{13}$
% ...
\\
\vspace{1.5 cm}
\footnotesize
1. Dipartimento di Fisica dell'Universit\`a di Bari and INFN, 70126 
Bari,  Italy \\
2. Dipartimento di Fisica dell'Universit\`a di Bologna and INFN, 
 40126 Bologna, Italy \\
3. Physics Department, Boston University, Boston, MA 02215, 
USA \\
4. California Institute of Technology, Pasadena, CA 91125, 
USA \\
5. Department of Physics, Drexel University, Philadelphia, 
PA 19104, USA \\
6. Laboratori Nazionali di Frascati dell'INFN, 00044 Frascati (Roma), 
Italy \\
7. Laboratori Nazionali del Gran Sasso dell'INFN, 67010 Assergi 
(L'Aquila),  Italy \\
8. Depts. of Physics and of Astronomy, Indiana University, 
Bloomington, IN 47405, USA \\
9. Dipartimento di Fisica dell'Universit\`a dell'Aquila  and INFN, 
 67100 L'Aquila,  Italy \\
10. Dipartimento di Fisica dell'Universit\`a di Lecce and INFN, 
 73100 Lecce,  Italy \\
11. Department of Physics, University of Michigan, Ann Arbor, 
MI 48109, USA \\	
12. Dipartimento di Fisica dell'Universit\`a di Napoli and INFN, 
 80125 Napoli,  Italy \\	
13. Dipartimento di Fisica dell'Universit\`a di Pisa and INFN, 
56010 Pisa,  Italy \\	
14. Dipartimento di Fisica dell'Universit\`a di Roma ``La Sapienza" and INFN, 
 00185 Roma,   Italy \\ 	
15. Physics Department, Texas A\&M University, College Station, 
TX 77843, USA \\	
16. Dipartimento di Fisica Sperimentale dell'Universit\`a di Torino and INFN,
 10125 Torino,  Italy \\	
$a$ Also Universit\`a della Basilicata, 85100 Potenza,  Italy \\
$b$ Also INFN Milano, 20133 Milano, Italy\\
$c$ Also Istituto TESRE/CNR, 40129 Bologna, Italy \\
$d$ Also Scuola Normale Superiore di Pisa, 56010 Pisa, Italy\\
$e$ Also  Faculty of Sciences, University Mohamed I, B.P. 424 Oujda, Morocco \\
$f$ Also Dipartimento di Fisica, Universit\`a di Roma Tre, Roma, Italy \\
$g$ Also Universit\`a di Trieste and INFN, 34100 Trieste, 
Italy \\
$h$ Also Dipartimento di Energetica, Universit\`a di Roma, 
 00185 Roma,  Italy \\
$i$ Also Institute for Nuclear Research, Russian Academy
of Science, 117312 Moscow, Russia \\
$l$ Also Institute for Space Sciences, 76900 Bucharest, Romania \\
$m$ Swarthmore College, Swarthmore, PA 19081, USA\\
$n$ RPD, PINSTECH, P.O. Nilore, Islamabad, Pakistan \\
$p$ Also INFN Catania, 95129 Catania, Italy
\end{center}

\normalsize
\begin{abstract}
An experimental  study of the production of up-going charged particles
in inelastic interactions of down-going underground muons is reported, using
data obtained from the MACRO detector at the Gran Sasso Laboratory. 
In a sample of $12.2\times 10^6$ single muons, corresponding to a detector
livetime of $1.55\ y$, 
243 events are observed having an up-going particle associated with
a down-going muon. These events are analysed to determine the range and emission
angle distributions of the up-going particle, corrected for detection and 
reconstruction efficiency.

Measurements of the muon neutrino flux by underground detectors are often based
on the observation of through-going and stopping muons produced in $\nu_\mu$
interactions in the rock below the detector.
Up-going particles produced by an undetected down-going muon 
are a potential background source in these measurements.
The implications of this background for neutrino studies using MACRO are 
discussed.
\end{abstract}
\vskip 1.0 truecm
\newpage

\section*{1. Introduction}

One of the primary goals of the MACRO detector at the Gran Sasso underground 
laboratory is the
measurement of the flux of atmospheric muon neutrinos.
In this paper we present the measurement of a potential background in these
studies; charged up-going particles (primarily pions)  produced in inelastic 
interactions of down-going muons in the rock
surrounding the detector. If the down-going muon is not detected 
 the up-going pion may be misidentified as resulting from
a neutrino-induced event. This will occur, for example,  when the $\mu$ passes 
near but not
through the detector. This type of event, therefore, represents a
background in the measurement of the flux of neutrino-induced upward 
through-going and stopping muons. 
The events under study, although similar in topology to muon backscattering, 
have an observed rate which is many orders of magnitude
larger than that expected from the large angle scattering of muons 
\cite{elbe,deep2}. The relatively low energy up-going
tracks are primarily charged pions produced in hard muon scattering on a nucleon 
$N$,
$\mu + N \rightarrow \mu + \pi^{\pm}+ X $.
Hadron production by underground muons
has been recently discussed \cite{neutron} in conjunction with the problem of 
neutron 
background in the measurement of the $\nu_e/\nu_\mu$ atmospheric neutrinos ratio 
by large water Cherenkov detectors \cite{emuratio}.
Hadron production by muon interactions in matter is also a source of background 
in 
radiochemical solar neutrino experiments \cite{solar}.

In this paper we present the results of a study of these events carried
out with the MACRO detector.
In Section 2 the MACRO apparatus and the method used
to identify muon-induced up-going charged particles
are described. The details of the analysis of the data are given in Section 3.
In Section 4 the observed range and emission angle distributions for the 
up-going
particle are presented.
In Section 5 the detector response and the tracking efficiency 
for these events are calculated. In Section 6 the results are interpreted within 
the
framework of a model of the photonuclear interactions of high energy muons
provided by the FLUKA generator.
Finally, in Section 7 the background due to
up-going particles produced by hard muon scattering 
in the measurement of neutrino-induced muons is calculated for the MACRO 
experiment.

\section*{2. Muon-induced Up-going Particles in MACRO.}

The MACRO detector, described in detail in Ref. \cite{MACRO93},
is located in Hall B of the Gran Sasso underground laboratory.
The detector has global dimensions of $76.6m \times\ 12m \times\ 9.3m$,
and is divided longitudinally in six similar supermodules and
vertically in a lower ($4.8\ m$ high) and an upper part
($4.5\ m$ high).
The active detectors include  horizontal and vertical planes 
of limited streamer tubes for
particle tracking, and liquid scintillation counters for fast timing.
In the lower part, the eight inner planes of limited streamer tubes
are separated by passive absorber (iron and rock $\sim 60 g\ cm^{-2}$)
in order to set a minimum threshold of $\sim 1\ GeV$ for vertical muons
crossing the detector. The upper part of the detector
is an open volume containing electronics and other equipment.
The horizontal streamer tube planes are instrumented with
external pick-up strips at an angle of $26.5^o$ with respect to the
streamer tube's wire, providing stereo readout of the detector hits.
The transit time of particles through the detector is obtained 
from the scintillation counters by measuring the mean time at which signals are
observed at the two ends of each counter, and taking the difference in the
measured mean time between upper and lower counters.    
The particle direction in the detector is given by the sign of this difference.
 
\begin{figure}[tb]
\centering
\psfig{figure=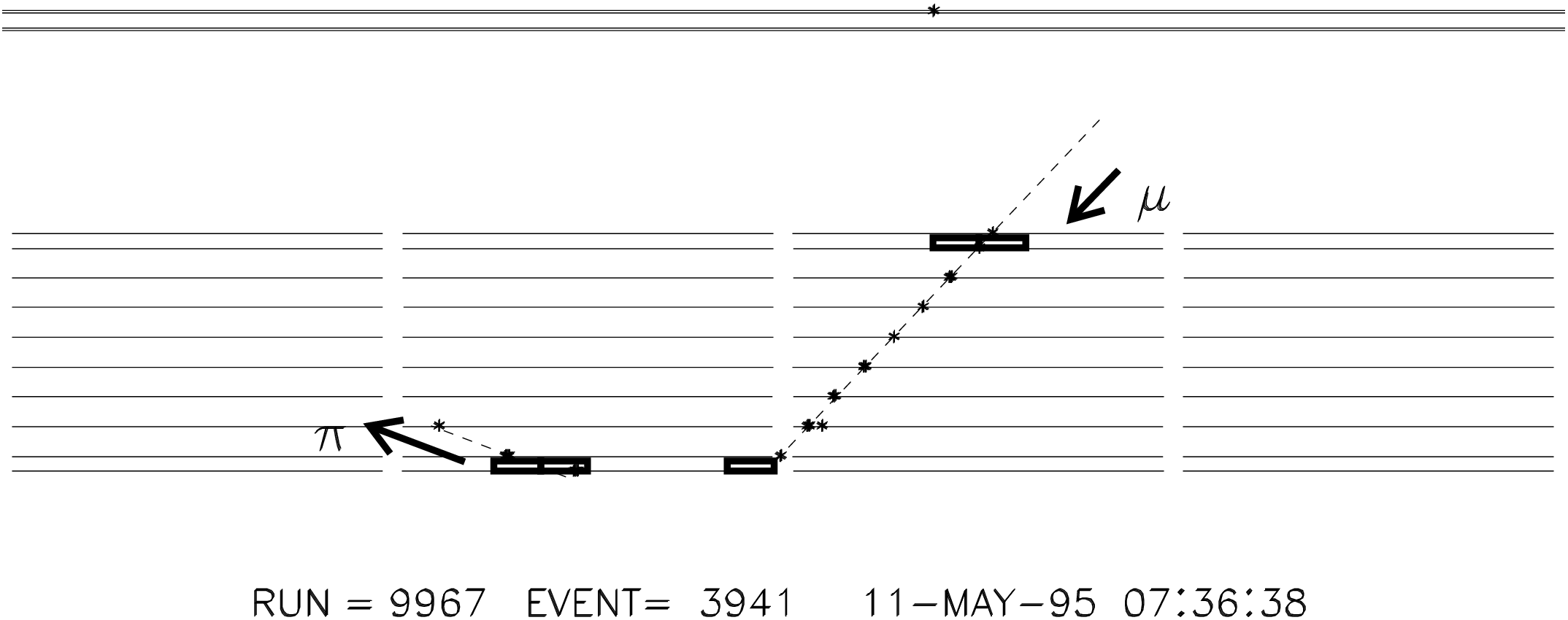,height=2.2in,%
bbllx=0bp,bblly=225bp,bburx=370bp,bbury=600bp}
\protect\caption[]
{\label{fig:inte2}\small
On-line display of a typical muon interaction in the rock below the apparatus
giving rise to an up-going charged particle. The rectangular boxes indicate 
scintillator 
counter hits and the points are streamer tube hits.
The tracks are shown in the wire view of the streamer tubes.
Only two supermodules are drawn.}
\end{figure}

The apparatus can observe charged particles 
produced at large angles by hard muon interactions if the reaction 
happens either inside the apparatus or in the rock and concrete
below the apparatus.
In the latter case, a secondary particle produced at a large angle in the 
reaction 
is seen as an up-going particle in the apparatus (see Fig. \ref{fig:inte2}). 
If the muon passes through the detector, such an event consists of two tracks 
converging somewhere below the detector, 
both in the wire and strip projective views of the tracking system.
The secondary particle reaches the scintillator counters with a 
positive time delay with respect to the counters hit by the muon.

In the sample of muon plus up-going pion events,
the down-going particle is usually the longer of the two tracks, and is 
reconstructed
by the standard MACRO tracking procedure \cite{MACRO93}.
This tracking algorithm is optimized for the reconstruction of single
and multiple muon events, and requires that at least four central streamer tube
planes are crossed by the particle, corresponding to a minimum of $\sim 200\ g\
cm^{-2}$ of detector grammage traversed.
For this study, a tracking algorithm was developed to reconstruct 
the short, diverging tracks characteristic of the events of interest.
This algorithm searches for alignments  between a cluster of scintillation
counters and at least two streamer tube hits in a 4 meter wide region
centered on the counter cluster.
The scintillator cluster is defined as a contiguous group of 
neighboring counters simultaneously hit by one or more particles.
The reconstructed coordinates of the scintillation cluster, obtained from the
identity of the hit counters and the timing difference between the two ends of
the counters, are used by the tracking algorithm as an additional track point.
A track candidate consists of an  alignment between at least three hits in a 4
meter wide region, and is referred to below as a short track.
The minimum vertical depth for events reconstructed by this algorithm is 
approximately
$25\ g\ cm^{-2}$.
\newpage

\section*{3. Event Selection}

\begin{figure}[tb]
\centering
\psfig{figure=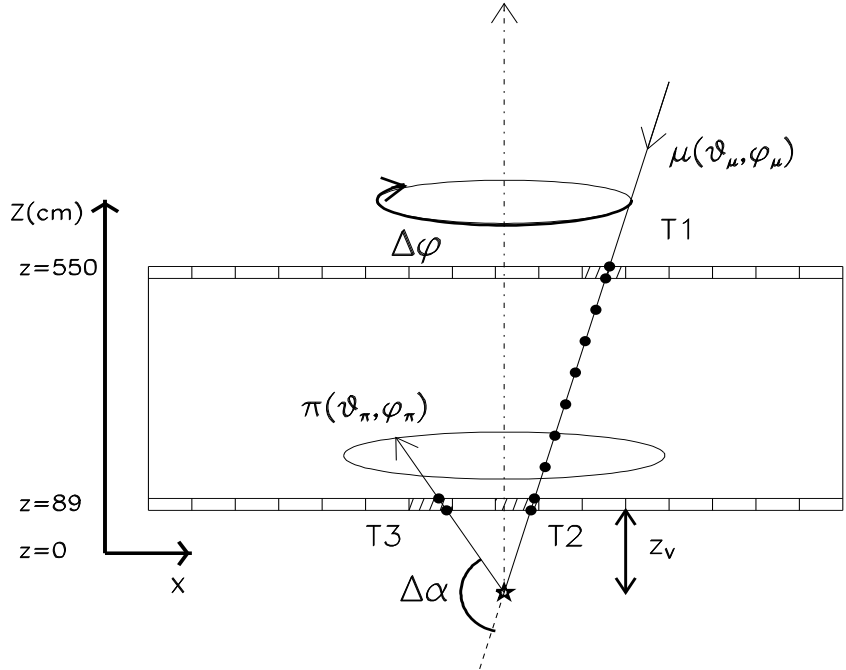,height=3.0in,%
bbllx=0bp,bblly=225bp,bburx=370bp,bbury=575bp}
\protect\caption[]
{\label{fig:doup}\small
Sketch of an event containing a down-going muon with direction
$(\vartheta_\mu,\varphi_\mu)$ producing an
up-going pion with direction $(\vartheta_\pi,\varphi_\pi)$.
The arrows indicate the direction of the particles,
deduced from the timing information obtained from the scintillation counters
($T_1,T_2$ and $T_3$). The filled points correspond to the streamer tube hits.
The zenith and azimuth emission angles $\Delta\alpha,\Delta\varphi$
of the pion with respect to the muon direction are indicated.}
\end{figure}

The data sample used in this study includes events 
in which a single muon is reconstructed by the standard tracking 
algorithm 
(see Fig. \ref{fig:doup}, for a sketch). The short track algorithm is then
used to search for events in which a secondary up-going pion enters the
detector from below and crosses the bottom layer of scintillator counters and
at least two  streamer tube planes  before coming to rest in the detector. 
Finally, the
timing information from the scintillation counters 
is used to identify the direction of the two particles.
As demonstrated below, the pion and the muon are resolved with high efficiency
if the wire hits belonging to the two tracks are separated by at least $\sim 75\ 
cm$.

Two different data periods have been analyzed. 
In the first (sample A, from  December 1992 through June 1993),
the upper part of the apparatus was not 
in acquisition.   In the second period 
(sample B, from April 1994 until January 1996)
the entire apparatus, including the upper part was in acquisition.
Table 1 contains the detector livetime,
the number of single downgoing muons,
the number of reconstructed short tracks, and the number of final muon plus 
up-going pion events
obtained in the two data taking periods. 
As shown in the table, a short track is reconstructed in 31717 of the
$12.2\times 10^6$ single muon events. 
These are referred to below as double track events.
A visual scan of a subsample of 15\% of these events 
has been performed to determine their nature.
In the most common case (76\% of events in sample A, and 69\% in sample B), 
the short track in the event belongs 
to a second downgoing muon, parallel to the first. 
The difference in the fraction of events of this type  between the two data
samples is due to the improved acceptance for double muon events during the 
period in which  
the entire apparatus was in acquisition.
Other topologies, all producing an incorrectly reconstructed short track,
are due to electromagnetic showers inside
the detector (17\% of events in the two periods), 
incorrect track point assignment (4\%), muon interactions
occurring inside the detector (5\%) and low momentum muons
undergoing significant multiple scattering in the detector (2\%). In all of 
these
event types the two reconstructed tracks do not possess a
common vertex point below the detector.

\begin{table}
\begin {center}
\begin{tabular}
{|c||c|c||c||c||}\hline
        & \ Single        & Double & Livetime & Down muon \\ 
Sample  &downgoing muons      & tracks  &  (h)   & + up pion \\ \hline
        &                     &         &        &         \\
      A & $3.3 \cdot {10^6}$ & $12216$ & $4235$ & $65$    \\ 
        &                     &         &        &         \\
      B & $8.9 \cdot {10^6}$ & $19501$ & $9351$ & $178$   \\ 
        &                     &         &        &         \\ \hline
   A+ B & $12.2 \cdot {10^6}$& $31717$ & $13586$& $243$   \\ 
        &                     &         &        &         \\ \hline
\end{tabular}
\end {center}
\caption {Number of single down-going muons and of
events with an additional short track (double tracks) in the two 
data periods A and B. In the fourth column the livetime $(h)$ of the apparatus
is shown. The number of identified muon plus up-going pion 
events is  given in the last column.}
\end{table}

The event topologies identified as background 
in the visual scan are rejected by 
applying the following analysis cuts to the sample of double track events.
Double muon events are rejected by requiring that the
direction of the two tracks differ by more than $26^o$.
Electromagnetic showers are rejected by requiring that the
total number of streamer tube hits not belonging to the muon track
be less than 20, while  muon interactions inside the apparatus and
low energy muons are rejected by requiring that the
intersection point of the two tracks be outside the detector volume.
3467/31717 events pass these cuts and
have been visually scanned to obtain the final data sample. Events 
eliminated in this final visual scan include in most cases multiple and/or
wrongly reconstructed short tracks, produced 
by a double muon event or by a muon electromagnetic shower.
As a result, 65 events in sample A and 178 in sample B 
are classified as muon plus up-going pion events.

The probability of rejecting a genuine muon plus up-going pion event through
the software cuts has been
evaluated by scanning a random subsample of 6200 double track events,
representing $\sim$20\% of the full double track event set, for muon plus 
up-going pion events.
54 events are found, of which 3 are rejected 
when the software cuts are applied. The corresponding selection efficiency is
$\epsilon_s=94\%$. An independent check of $\epsilon_s$
is obtained by applying the software cuts to  the sample 
 of simulated (discussed below) down-going muons plus up-going pions; 
151/3347=4.5\% of 
the simulated events are discarded.
Based on the number of events in the final data sample, and a selection
efficiency  $\epsilon_s\ = 94\%$,
the rate of detected up-going pions 
per down-going muon at the MACRO depth is:
$$n_{\pi/\mu}= {{243\over 0.94 \times 12.2\ 10^6}} = 
(2.1\pm 0.2)\ 10^{-5}\eqno(1)$$

\section*{4. Analysis}

The 243 events in the final data sample are now used to obtain
the distribution of the range of the up-going pion, the angle between the
$\mu$ and $\pi$, and the radial distance $D$ between the pion and the 
muon at a vertical position corresponding to the center of the bottom layer of 
scintillation counters.

Referring to Fig. \ref{fig:doup}, the point at which the muon
interaction occurs, $\vec{x}_v =(x_v,y_v,z_v)$, is defined as 
the intersection of the observed muon and pion tracks. 
The vertex depth $z_v$ is obtained separately in the wire ($z_v^w$) and strip
($z_v^s$) projection.
In 75 of the 243 events in the final data set the difference between the vertex
depth in the wire and strip views exceeds 50 cm; this event subsample is
defined as poorly reconstructed in the following.

\begin{figure}[tb]
\centering
\psfig{figure=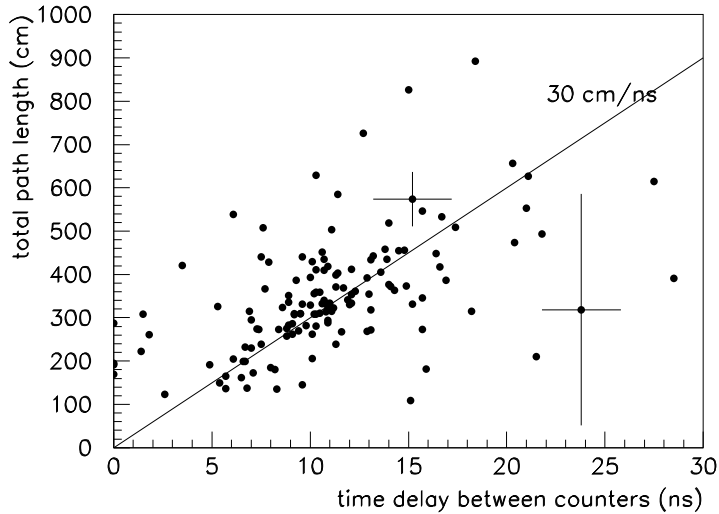,height=2.5in,%
bbllx=0bp,bblly=225bp,bburx=240bp,bbury=475bp}
\protect\caption[]
{\label{fig:tvsl}\small
Total path length (see text) versus time delay $\Delta t = (T_3 - T_2)$ 
between scintillation counter hits in the bottom plane.
Only the 168 events for which $|z_v^w - z_v^s| <\ 50\ cm$ 
are included. The large spread around the central line is due to the combined
uncertainty in the total path length and timing reconstruction. As an example,
error bars are drawn for two events.
For three events, $T_2$ is missing and  $\Delta t=0$}
\end{figure}

The timing information obtained from the scintillation counter system is  used
to  determine the direction of motion of each particle, and as an independent
check of the vertex location obtained from the tracking system.
Using the convention of Fig. \ref{fig:doup}, a
down-going particle will have a transit time difference  $(T_2 - T_1) >0$. The 
bottom counter
crossed by the short track is reached at a time $T_3$,  with
$(T_3 - T_2) >0$. 
For 206 of the 243 events in the final data sample the requirement $(T_3 - 
T_2)>0$ is fulfilled. 
For the remaining 37 events, either the down-going muon 
misses the counters of the bottom scintillation plane ($T_2$ not present)
or the two particles intersect scintillation counters in 
supermodules read out by different data acquisition systems
(for a small fraction of the time, the electronics
to synchronize the information between the supermodule readout 
systems was non-functional).
Because the timing error seen in these events can be accounted for, all 37 are
classified as muon plus up-going pion events and are included in the subsequent 
analysis.

The total path length between the two scintillation counter hits, given by
the path length of the muon from the scintillator $T_2$ to  $\vec{x}_v $  plus 
the path length
of the pion from   $\vec{x}_v $  to the scintillator $T_3$, is determined for
events in the final data sample which satisfy the condition  $|z_v^w - z_v^s| <\ 
50\ cm$. 
The correlation between the total path length and the time delay $(T_3 - T_2)$ 
is shown in Fig. \ref{fig:tvsl}. The linear correlation seen in this figure is 
consistent 
with the hypothesis that the up-going particle is produced at the vertex point.
The uncertainty in the total path length depends on the uncertainty in the
vertex location, and ranges from $50\ cm$ up to $250\ cm$ for the worst.
The average
uncertainty in the time delay between the two scintillation counters is
approximately 2 ns for this data set.  

\begin{figure}[tb]
\centering
\psfig{figure=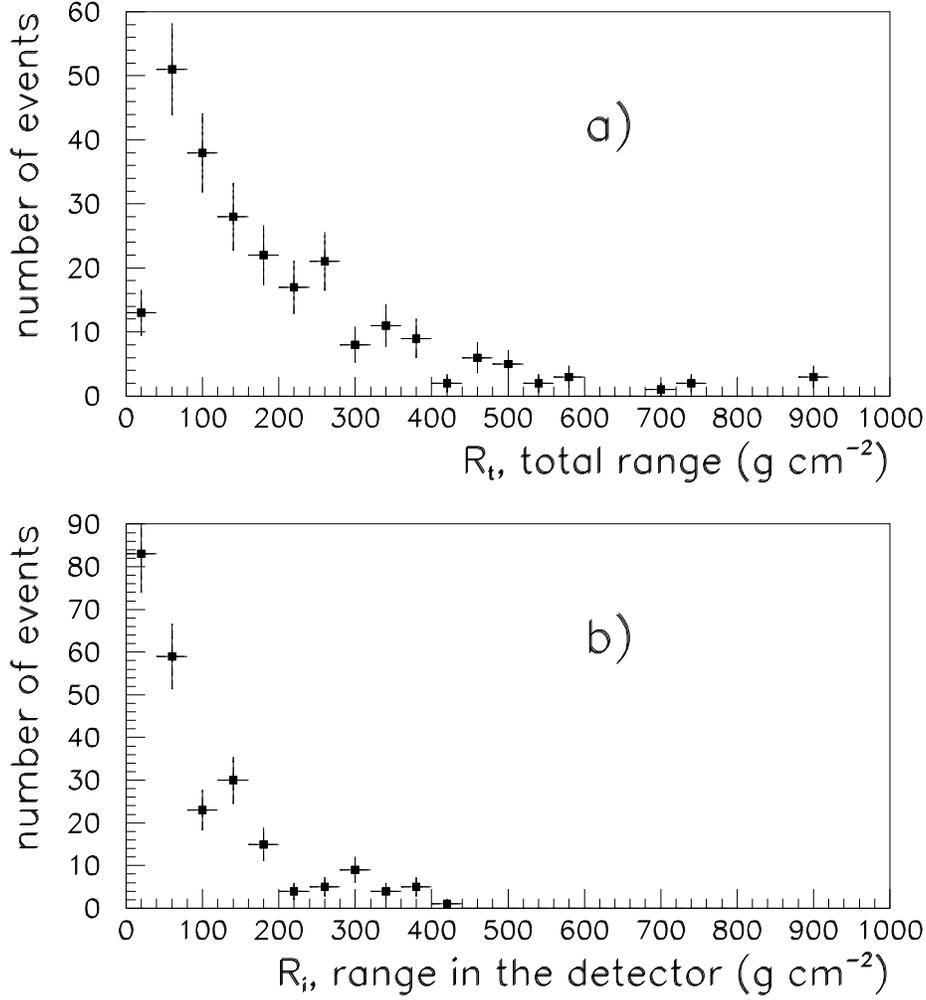,height=4.5in,%
bbllx=0bp,bblly=225bp,bburx=370bp,bbury=575bp}
\protect\caption[]
{\label{fig:grcm}\small
a) Distribution of the reconstructed total range $R_t$
of the up-going particles for which $|z_v^w - z_v^s| < 50\ cm$. 
Events with a poorly reconstructed vertex are excluded
from this distribution because of the large uncertainty in $R_t$ for these 
events. 
b) Distribution of the grammage traversed in the detector,  $R_i$,
for all the 243 up-going particles in the final data set.}
\end{figure}

\begin{figure}[tb]
\centering
\psfig{figure=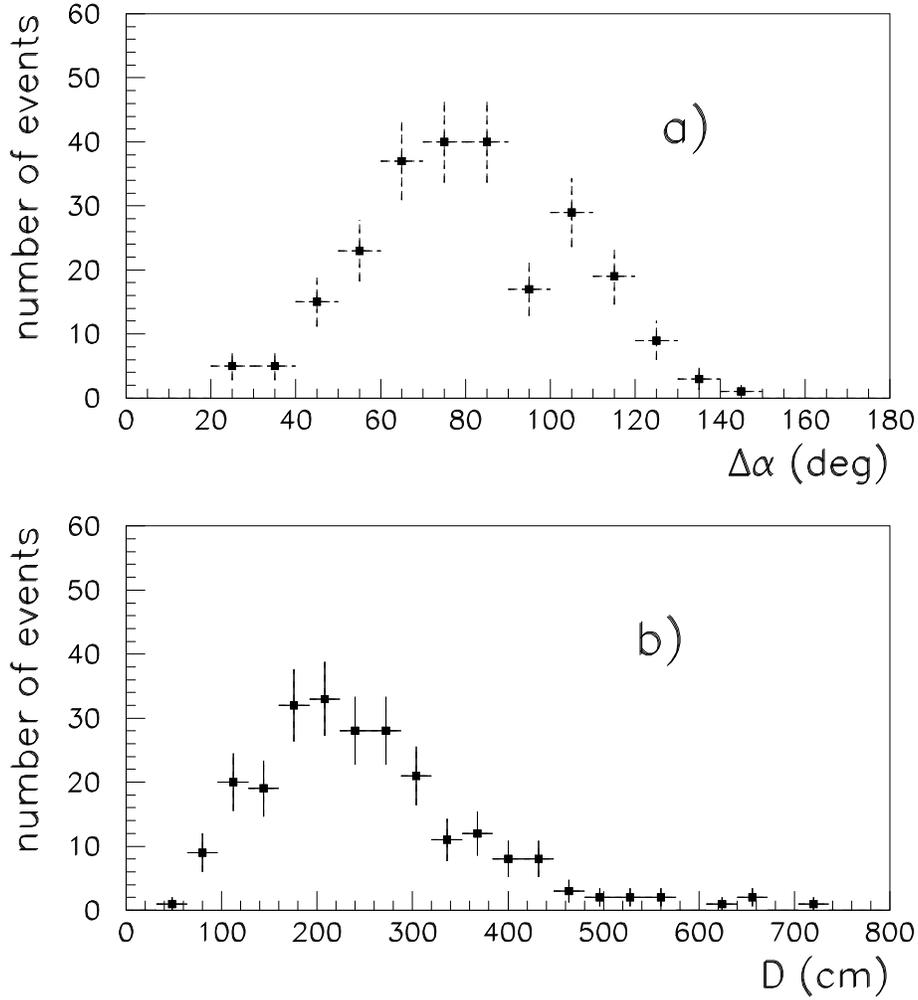,height=4.5in,%
bbllx=0bp,bblly=225bp,bburx=370bp,bbury=575bp}
\protect\caption[]
{\label{fig:alfad}\small
a) Measured distribution of the $\pi - \mu$ scattering angle 
$\Delta\alpha$ and b)
of the distance $D$ between the down-going muon and the
up-going pion at the $z$ location of the center
of the bottom layer of scintillation counters. }
\end{figure}

The range of the up-going particle,  $R_t$, is defined  as the distance between
the interaction point ($\vec{x}_v $) and 
the position of the uppermost streamer tube hit in the track, ($\vec{x}_s$).
The range obtained in this way is, in most cases, 
an underestimate of the true range of
the pion, by an amount as large as the vertical depth of the
concrete absorber layer between streamer tube planes, $60\ g\ cm^{-2}$.
The particle range is calculated by dividing the path length from $\vec{x}_v $ 
to $\vec{x}_s$
into 200 steps and considering the density of the material at each step.
The sum of the product of the density times the step length
gives the total range in $g\ cm^{-2}$.
This procedure makes use of a detailed description of the distribution
and density of the material inside and around the apparatus.
In Fig. \ref{fig:grcm}a the distribution of the 
total range of the pion, $R_t$, is shown.
Only muon plus up-going pion events satisfying the condition  $|z_v^w - z_v^s|
< 50\ cm$ are included in this distribution. 
The partial range, $R_i$, of the  up-going particle
inside the apparatus is also calculated, and is shown
in Fig. \ref{fig:grcm}b.
$R_i$ is calculated from the bottom of the detector
to the last active streamer tube hit in the up-going track.
All the 243 muon plus up-going pion events are included 
in  Fig. \ref{fig:grcm}b, since the
position of the vertex is not used in the calculation of $R_i$. 
It can be seen from these figures that a significant fraction of
these events contain an up-going particle which penetrates enough 
material to be incorrectly identified as an up-going muon when the down-going 
particle is undetected.

Fig. \ref{fig:alfad}a shows the measured
distribution of the pion emission angle, $\Delta\alpha$, while
Fig. \ref{fig:alfad}b shows the distribution of the distance $D$
between the muon and the up-going pion at a vertical position corresponding to 
the center of 
the bottom layer of the scintillator counters.
The distributions, as discussed in the next section, decrease for small
$D$ as a result of a low event reconstruction efficiency, 
and for small $\Delta\alpha$ as a consequence of the low muon rate
at large zenith angle.
The radial distance between the muon and the up-going pion is less than $4\ m$,
corresponding to roughly  one third of the lateral size of one MACRO
supermodule, in 90\% of the events, as discussed in Section 7.
\newpage

\section*{5. Reconstruction Efficiency}

\begin{figure}[tb]
\centering
\psfig{figure=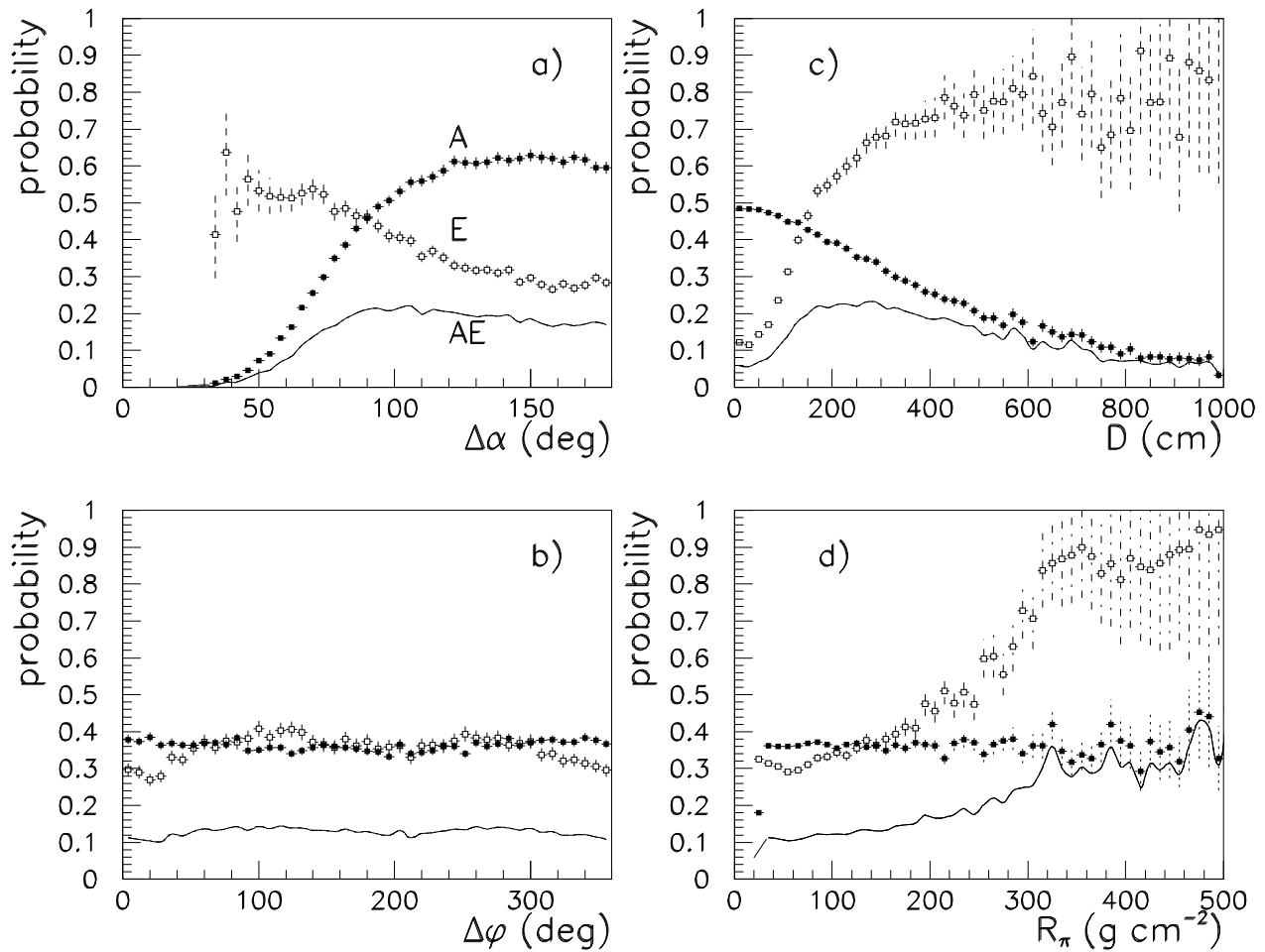,height=4.5in,%
bbllx=0bp,bblly=225bp,bburx=370bp,bbury=575bp}
\protect\caption[]
{\label{fig:para}\small
Distribution of the 
probability $A$ for events to be accepted, of the tracking efficiency
$E$, and of the reconstruction efficiency $AE$ versus
$(a)\ \Delta \alpha\ ,(b)\ \Delta \varphi\ ,\ (c)\  D$, and $ (d)\ R_\pi$.
Each quantity is averaged over the three remaining variables.
Full points: $A$; empty points: $E$; solid curves: $AE$.}
\end{figure}

The probability of detecting an up-going pion produced  by a down-going muon
depends on the emission angles of the pion with respect to the muon, the pion
energy, and the vertex depth.  To determine the detection probability for these
events in MACRO, a sample of $5.0\times  10^5$ simulated muon plus up-going pion 
events
is distributed uniformly over cells in the parameter space 
${\cal P}$ defined by the variables $\Delta\alpha,\Delta\varphi,z_v$ and 
the pion range $R_{\pi}$ (see Fig.\ref{fig:doup}). 
The zenith angle $\Delta\alpha$ of the pion with respect to the
incoming muon direction is constrained to lie in the range
$0^o<\Delta\alpha <180^o$.
The vertex location $\vec{x}_v$ is taken at random locations below 
the detector, with the vertex depth at a distance 
$5<d_v <205 \ cm $ from the bottom layer. The  pion range inside the 
lower detector 
is chosen between $0<R_\pi<1000\ g\ cm^{-2}$. This corresponds 
to a value of the pion momentum at the detector surface, $p_\pi$, in the 
interval $0.1<p_\pi <2.0\ GeV/c $.
In each simulated event, the down-going muon direction 
$(\vartheta_\mu,\varphi_\mu)$ is extracted from the measured distribution of
atmospheric single muons.  The point at which the muon enters the detector is
distributed uniformly over the detector surface.
The pion direction, $(\vartheta_\pi,\varphi_\pi)$, is given by  $\vartheta_\pi = 
180^0-
\vartheta_\mu - \Delta\alpha$ and $\varphi_\pi = \varphi_\mu - \Delta \varphi$.
Using the events uniformly distributed on respect these
parameters, and the actual angular distribution of
the muons at MACRO depth, the reconstruction efficiency,
averaged on the muon directions, has been obtained
as a function of the chosen parameter set.

GMACRO \cite{MACRO93}, a GEANT-based \cite{GEANT} Monte Carlo of the
MACRO detector, is used to model the response to these particles. It
describes the experimental apparatus ({\it i.e.}
geometry, density, and the detector
response) in all its details. The propagation of low-energy pions is potentially 
sensitive to the treatment of charge-exchange and in-flight absorption
reactions. For this reason, the propagation of pions in the detector using the 
GEANT-GHEISHA (G-G)
hadronic interface has been compared with results using the GEANT-FLUKA (G-F)
interface (see \cite{GEANT} for references). We find that the two models
give essentially the same detection probability for muon plus up-going pion 
events.  For
example, G-F predicts that a $500\ MeV$ 
pion entering the detector from the bottom layer has a 
$20\%$ probability of a range within the detector smaller than $25\ 
g\ cm^{-2}$, while
G-G predicts a probability of $23\%$.  This range corresponds to the minimum 
distance
required to identify the up-going pion, and as a consequence, the 
number of accepted events does not differ significantly
for the two cases. On the other hand, the energy distribution of the up-going 
pions, 
unfolded from the measurement of the range, is quite different for the two
models. This can be illustrated by noting that
the same $ 500\ MeV$ pion has a probability of $57\%$ of having a range smaller 
than 
$100\ g\ cm^{-2}$ in G-F, while G-G predicts a $74\%$ probability. In the 
following, we report the results obtained with the GEANT-FLUKA simulation.

\begin{figure}[tb]
\centering
\psfig{figure=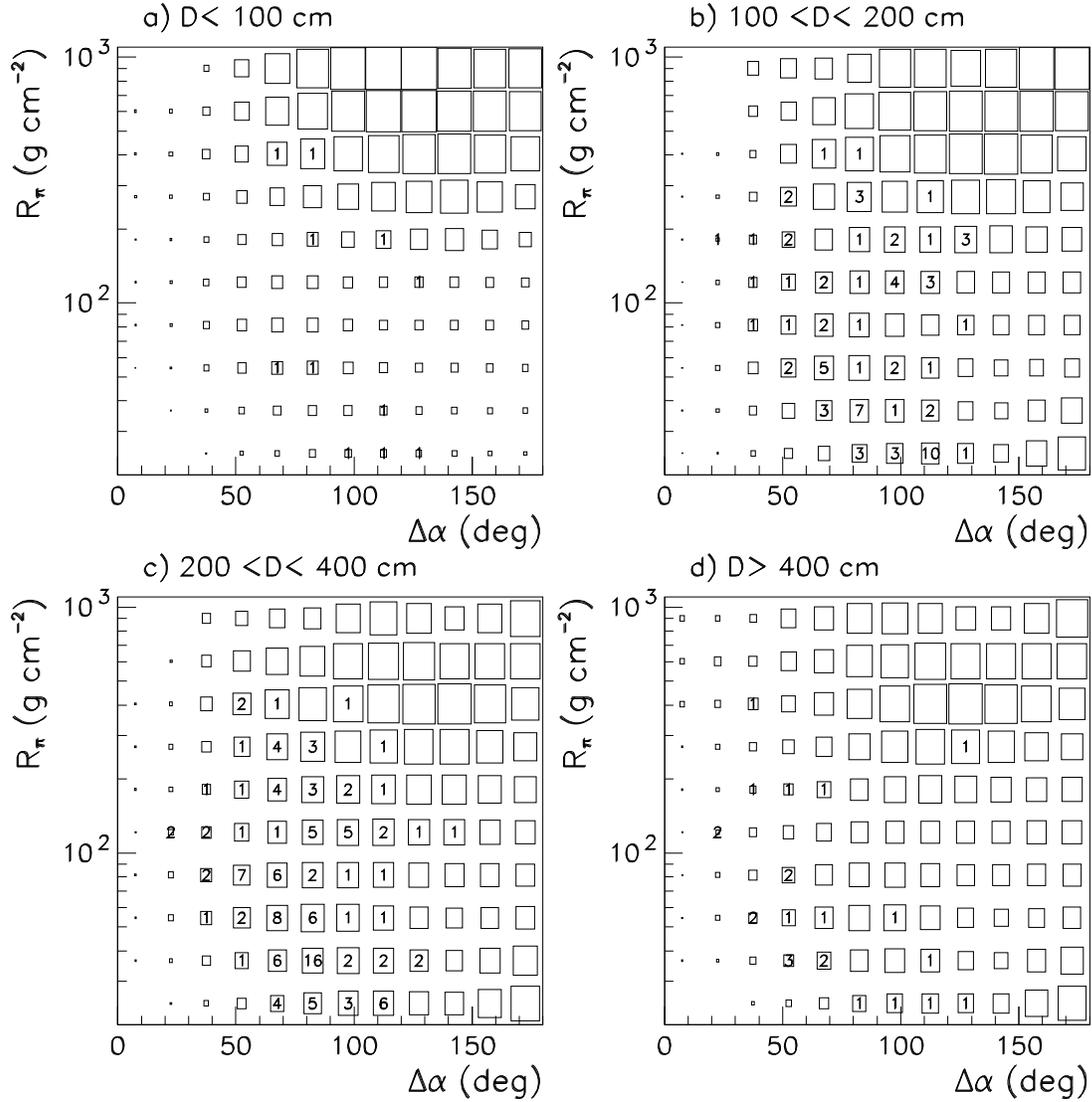,height=4.8in,%
bbllx=0bp,bblly=225bp,bburx=370bp,bbury=575bp}
\protect\caption[]
{\label{fig:boxes}\small
Reconstruction efficiency $AE$ in the ($R_\pi,\Delta\alpha,D$) 
parameter space, for 4 intervals of the distance $D$,
12 bins of $\Delta\alpha$ and 10 bins of the 
measured range of the pion inside the detector $R_\pi$.
The computed values of $AE$ are proportional to
the box area. For example, the upper row in (d)
corresponds to values of $(AE \cdot 100)$ equal to
$1,1,3,18,29,48,52,47,40,30,48$ and $76$. In each cell the number of detected
muon plus up-going pion events at the given ($R_\pi,\Delta\alpha,D$) values is 
indicated.}
\end{figure}

The result of processing the simulated muon plus
up-going pion events in GMACRO is an output data set whose format is
essentially identical to that of real data.
This data set has been processed with the analysis chain used for the 
real data, described in section 4.
A simulated event is defined as {\it accepted},
if the down-going muon is tracked and if the pion enters the detector and 
crosses the lower two horizontal streamer tube planes 
and the horizontal scintillator layer in between.
The minimum amount of material traversed by a pion satisfying this requirement 
is $\sim 25\ g\ cm^{-2}$.
The event is defined as {\it tracked} if the pion track is reconstructed, as
described in Section 2.
The acceptance probability and
tracking efficiency for the simulated events is now obtained for each cell in
the parameter space ${\cal P} = (R_{\pi},\Delta\alpha,\Delta\varphi,
D(z_v))$.
The radial distance $D$ is uniquely
determined for a given pion and muon directions $(\vartheta_\pi,\varphi_\pi)$, 
$(\vartheta_\mu,\varphi_\mu)$ and vertex depth $z_v$.
The acceptance probability for an event is defined as
$$ A(R_{\pi},\Delta\alpha,\Delta\varphi,D)
 = {{\:number\:of\:accepted\:event} \over 
{number\:of\:simulated\:events}}\eqno(2)$$
while the tracking efficiency  is defined as
$$ E(R_{\pi},\Delta\alpha,\Delta\varphi,D)
 = {{\:number\:of\:tracked\:events} \over
{number\:of\:accepted\:events}}\eqno(3)$$

Fig. \ref{fig:para} shows the acceptance probability $A$, 
the tracking efficiency $E$, and the reconstruction efficiency $AE$ 
as a function of $\Delta \alpha\ ,\:\Delta \varphi\ ,\: D\,$ and $R_\pi$. 
In each case, the values of $A$ and $E$ are obtained 
by averaging over the remaining three variables.
Referring to Fig. \ref{fig:para}, 
very few pions emitted with a small  $\Delta \alpha$ (Fig. \ref{fig:para}a) 
are detected, due to the fact that these pions must be produced by 
a near-horizontal muon, and the integrated muon flux with $\vartheta_\mu>75^o$ 
is $6\times  10^{-3}$ of the total. 
In  Fig. \ref{fig:para}c, the decrease in the tracking efficiency for small $D$
is a result of the merging of the  pion and the muon tracks.  
Finally, the reconstruction efficiency 
increases as a function of the pion range within the detector
(Fig. \ref{fig:para}d), because the tracking efficiency increases with the
number of streamer tube hits in the track.
Since (Fig. \ref{fig:para}b) $ AE $ is essentially independent of the azimuthal
scattering angle $\Delta \varphi$, it is assumed in the following that $ AE = 
AE(R_{\pi},\Delta\alpha,D) $.

\begin{figure}[tb]
\centering
\psfig{figure=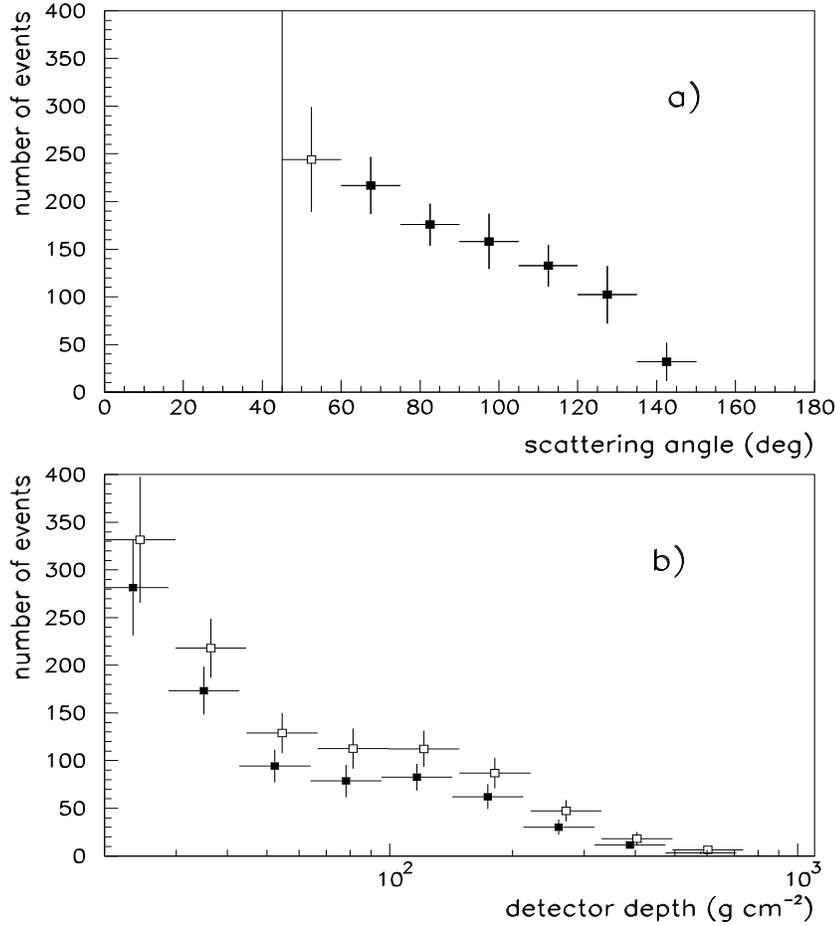,height=4.0in,%
bbllx=0bp,bblly=225bp,bburx=370bp,bbury=575bp}
\protect\caption[]
{\label{fig:unfold}\small
a) Detector-unfolded distribution of the zenith scattering angle $\Delta\alpha$ 
of the pion with respect to the muon direction, for $\Delta\alpha>45^o$.
b) Detector-unfolded distribution of the up-going pion range inside the 
detector. Open points: pions with $\Delta\alpha>45^o$;
full points: pions with $\Delta\alpha>60^o$ (b). Only statistical 
errors are shown. The number of events corresponds to $1.55\ y$ of data taking.}
\end{figure}

In Fig. \ref{fig:boxes}, the value of  $AE$ is shown as a function of 
$R_\pi$ and $\Delta\alpha$ in four intervals of the distance $D$:
(a) $D<100\ cm$, (b) $100<D<200\ cm$, (c) $200<D<400\ cm$ and (d)
$D>400\ cm$.
The box sizes are proportional to the reconstruction efficiency $AE$
for detecting an event at a given ($R_\pi$,$\Delta\alpha$,$D$). 
In the same figure, the numbers inside the boxes correspond to the 
distribution of real events.
In the four intervals of $D$ there are 11, 78, 130 and 24 detected 
events with a muon plus an up-going pion, respectively.
The number of detected events $N_{det}(i,j,k)$ in each cell 
of Fig. \ref{fig:boxes}
can be considered as the convolution (plus statistical fluctuations) of the
reconstruction efficiency $AE(i,j,k)$ with the 
unknown physical distribution of the
events as a function of the variables in the 
parameter space ${\cal P}$.
To obtain these physical distributions, we evaluated the corrected
number of events as the number of detected events in each cell 
divided by the reconstruction efficiency: 
$N_{corr}(i,j,k) = N_{det}(i,j,k) / AE(i,j,k)$.
Because of the uncertainties in the reconstruction of $\Delta\alpha$ and
$R_\pi$ from the measured track parameters, there is a small probability
that an event in the $(l,m,n)$ cell of ($R_\pi,\Delta\alpha,D$) 
will be measured in the $(i,j,k)$ cell.
This probability, which results in a  smearing of the
$N_{det}(i,j,k)$ distribution, has been evaluated using the simulated sample.
The mean detection efficiency for $\Delta\alpha>45^o$,
averaged over the other parameters of ${\cal P}$, 
is found to be 
$$\epsilon_r =  \sum_{i,j,k} N_{det}(i,j,k) / \sum_{i,j,k} N_{corr}(i,j,k)
= (21 \pm 2_{stat} \pm 4_{sys})\%\eqno(4)$$
The systematic uncertainty is estimated by examining the effect of changing the 
sizes of the 
$(R_\pi,\Delta\alpha,D$) cells, and by evaluating the difference in the results 
obtained using
the two hadronic interfaces to GEANT.

\begin{figure}[tb]
\centering
\psfig{figure=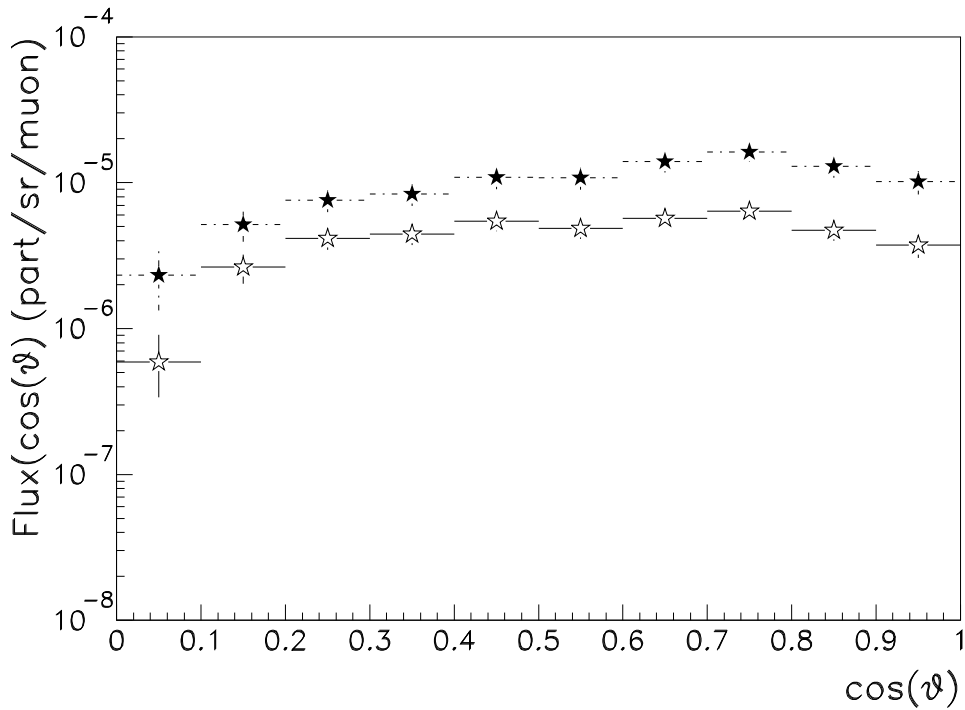,height=3.5in,%
bbllx=0bp,bblly=225bp,bburx=370bp,bbury=575bp}
\protect\caption[]
{\label{fig:cosze}\small
Angular distribution of the charged particles emerging
from the floor, $\vartheta$, relative to the vertical direction.
The flux is normalized to the total number of detected single muons
($12.2\times 10^6$). Open stars: uncorrected data. Full stars:
corrected and detector-unfolded data for 
pions with scattering angles $\Delta\alpha > 45^o$. The pion energy
threshold is around $100\ MeV$. Only statistical errors are shown.}
\end{figure}

From the distribution $N_{corr}(i,j,k)$, we obtain 
the detector-unfolded distribution of the pion-muon zenith scattering 
angle $\Delta\alpha$ and of the pion range inside the detector
$R_\pi$.
The two distributions, which correspond to the $x$ and $y$ projection of 
$\sum_k N_{corr}(i,j,k)$, respectively, 
are shown in Fig. \ref{fig:unfold}. Given the
low probability for events to be detected
at small pion emission angles, the
$\Delta\alpha$ distribution is shown for $\Delta\alpha>45^o$. 

Fig. \ref{fig:cosze} shows the 
angular distribution of up-going charged particles versus
$cos(\vartheta)$, where $\vartheta$ is the zenith angle in the
detector frame. The distribution is normalized to the
total number of down-going single muons ($12.2\times 10^6$), and is corrected
for reconstruction efficiency.

Using the detected number of muon plus up-going pion events and the mean 
detection efficiency quoted above, the estimated number of up-going 
pions emerging from the floor at the MACRO depth 
with angle of scattering  $\Delta\alpha>45^o$ with respect to the 
muon direction and range $> 25\ g\ cm^{-2}$, per down-going muon 
is found to be
$$N_{\pi/\mu}= {{ n_{\pi/\mu}\over \epsilon_r}} = 
(10^{+4}_{-2.5})\ 10^{-5}\eqno(5)$$

\section*{6. A Model of High Energy Muon Interactions}

To compare with the results of the previous section, 
estimates of pion production by underground muons
have been made using the hadronic interaction generator FLUKA\cite{fluka}
combined with a model of hard muon scattering
with the muon photonuclear cross sections of Ref. \cite{bezrukov}. 
The implications of this process for muon energy loss are discussed in 
\cite{lohman}.
The muon photonuclear cross section for the average muon energy
at the MACRO depth is
$\sigma_{\gamma}(\overline E_\mu \sim 280\ GeV)\simeq 0.40\ mb$.
This corresponds to a probability of $0.3\%$ for a $300\ GeV$ muon to
have one hard scattering in one meter of rock.
$\sigma_{\gamma}$ does not depend strongly on energy. For example,
$\sigma_{\gamma}(E_\mu=1000\ GeV)\simeq 0.55\ mb$.

In \cite{bezrukov} the interaction cross section for a muon of
a fixed energy is given in terms of the fractional energy loss and of the $q^2$ 
transferred.
Hadron production by muons is computed in the framework of
the Weizs\"{a}cker--Williams approximation \cite{weinz} 
for the radiation of virtual photons, and the vector dominance model.  Measured
photon-vector meson couplings are used. 
Some uncertainties and limitations of the simulation arise from the fact that
deep inelastic scattering is not included in FLUKA.
However, this process is dominated by low $q^2$ interactions.
In addition, although bremsstrahlung, pair production and other 
electromagnetic interactions of muons are considered,
the transport of secondary $e^+e^-$ and $\gamma$ is not performed. Therefore,
electromagnetic showers are not produced, nor is photo-production by real
photons considered. 

The differential energy 
spectrum $G(E_\mu,h)$ of muons at the MACRO depth, along with the
interaction model discussed above, have been used
to estimate the flux of hadrons produced by muons at MACRO.
An analytic approximation of $G(E_\mu,h)$ \cite{Gaisser} is used, which 
assumes that  the atmospheric inclusive muon spectrum above 1 TeV is described
by the power law $\Phi(E_\mu) = K E_\mu^{-\gamma}$,and gives a differential  
muon spectrum underground at the effective average depth $h$ of
$G(E_\mu,h) = K e^{-\beta h \left( \gamma-1\right) }
\left( E_\mu + \epsilon\left( 1-e^{-\beta h}\right) \right)^{-\gamma}$, with 
$\gamma$=3.7, $\beta$=0.4 $(km~w.e.)^{-1}$
and $\epsilon$=540 GeV \cite{Gaisser}. The average value of the
rock overburden for MACRO, weighted by the measured muon flux, is $h=3.8\ 
km.w.e.$.
Due to the shape of the rock overburden, there is a 
correlation between residual energy and angle. This correlation is neglected in
the model.

To simulate the production of up-going pions under MACRO, the muon energy is
extracted from the energy spectrum shown above in the energy range 
$1\ GeV < E < 10\ TeV$. An interaction of the muon in the rock floor
is then modeled,
and the energy spectrum of the up-going particles exiting the floor is 
determined. 
All  possible muon interactions are allowed, and the full hadronic 
cascade development is considered.
For each particle type, the yield as a function of the kinetic energy
and emission angle is then determined.
We find that, as expected,  the charged particle yield 
is dominated by charged pions, and that the differential pion spectrum 
decreases sharply with the kinetic energy 
and the emission angle $\Delta\alpha$.
 The charged kaon contribution is found to be 
lower than the pion contribution by about one order of magnitude. Protons 
(and neutrons) have an energy 
spectrum which is considerably softer than that of pions.

Next, in order to determine the number of muon plus up-going pion events
due to this process observed in MACRO, muons with the energy distribution shown
above are generated in $(45\times 180)$ cells of 
$(\vartheta_\mu,\varphi_\mu)$ 
using the intensity/depth relation given in
\cite{lama}, and  the slant depth
obtained using a rock map of the Gran Sasso 
mountain (see \cite{lama}).
The measured density of the Gran Sasso rock ($\rho = 2.657\ g/cm^3$) is used 
in place of the usual standard rock.
The absolute muon flux observed by MACRO provides the overall normalization
for the simulation.
Each simulated atmospheric muon undergoes a hard scattering in the rock
around the apparatus. Only pion production is considered, 
according to the differential pion yield $Y_{\pi/\mu}(E_\pi,\Delta\alpha )$
obtained  by the simulation procedure above.
The generated pions are then propagated to the apparatus. 
A simulated sample equivalent to $1.2\times 10^8$ detected single muons 
was generated, corresponding  to $N_y=14.4\ y$ of full detector
livetime.

Taking into account the real probability for hard scattering 
by an atmospheric muon, the predicted rate of muon plus up-going pion events
in which both particles enter in the detector is
$(8\pm 3)\ 10^{-5}\ \pi/\mu$. The quoted error includes the uncertainty in the
muon flux, the pion yield and the pion propagation in the detector. This
result is in agreement with the measured value, quoted in Section 5.

\section*{7. Evaluation of the Background due to Up-going Pions in Neutrino 
Studies.}

The flux of atmospheric muon neutrinos $\Phi_{\nu(\mu)}$
is often inferred from a measurement of the flux of up-going muons.
MACRO can measure $\Phi_{\nu(\mu)}$ in the $100\ GeV$ region
using through-going muons and, in the $5-10\ GeV$ region, using stopping muons.
An up-going pion induced by an interaction of a down-going atmospheric muon
in the rock around the detector results in a background event in
the neutrino measurement if the muon is undetected. 

The rate of events in which the down-going muon 
is not detected,  and in which the up-going pion  
simulates a neutrino-induced muon, has been evaluated using the simulation
described in the previous section. 
In the simulated sample, equivalent to $14.4\ y $ of data taking,
$63$ events are seen in which the up-going pion crosses two
scintillator layers and the muon is undetected.
Of these, $31\ $ events contain pions which traverse more than  $200\ g\ 
cm^{-2}$ of
detector material. These events satisfy all of the criteria \cite{upmu}
for an upward through-going muon. This gives a background rate due to up-going 
pions of
$$Upward\ throughgoing\ 
\mu\ background\ =\ (2.1 \pm 0.4_{stat} \pm 0.7_{sys})\quad events/year$$
\noindent The expected number of detected up-going muons from $\nu_\mu$
interactions in the rock below MACRO is approximately $200\
events/year$\cite{upmu}, giving a contamination from this background of 
approximately $1\%$.
The distribution of the cosine of the zenith angle of these background events is 
essentially flat.

Partially contained muon neutrino interactions in the apparatus,
giving rise to a down-going muon, and
up-going stopping muons induced by a $\nu_\mu$ interaction in the rock below the
apparatus,  are identified by applying appropriate constraints to the observed 
tracks
\cite{stop}. In the following, these events are called {\it stopping muons}.
The background for the stopping muon  sample is evaluated
by applying the analysis of ref. \cite{stop} to the simulated events.
127 events are found in which the up-going pion satisfies the criteria used in 
this
analysis for the identification of stopping muons, and  in which the muon
is not observed. This corresponds to a rate of these background events of
$$Upgoing\ stopping\ \mu\ background\  =\
(8.8 \pm 0.9_{stat} \pm 0.8_{sys})\ events/year$$ 
\noindent The expected rate of stopping muons 
is about $80\ events/year$ in MACRO, implying a contamination by up-going pions
in this sample of about $10\%$.

The background rates obtained above are specific to the MACRO detector.
For smaller detectors, the probability of detecting the pion while  missing the 
muon
increases.
For example, for a single MACRO supermodule 
the background/signal ratio for the
through-going sample is almost twice the value quoted above.
Other underground experiments measuring neutrino-induced muons
should have a similar contamination due 
to up-going pions, with the exact background
rate depending on the geometrical configuration 
of the detector, its depth, 
and on the momentum threshold applied to the up-going muons.
To our knowledge,  no experiment measuring the flux of atmospheric
neutrinos using up-going muons (see references in \cite{upmu}) has taken
into account this background. In particular, shallower experiments
for similar energy threshold
should have a significant contamination from these events, because 
the average muon energy, $\overline E_\mu(h)$, and the corresponding 
pion yield,
$Y_{\pi/\mu}$, decreases slowly with decreasing depth, while the total muon 
flux increases exponentially.

\section*{8. Conclusion}

In this paper we report the first measurement 
of up-going particles associated with down-going atmospheric muons
in an underground detector. 
This measurement was possible because of the large area, good tracking and good
time measurement capabilities of the MACRO apparatus.

In $12.2\times 10^6$ downgoing single muons we
find a total of 243 events 
with an identifiable up-going particle spatially and temporally
associated with a down-going muon.
We have analyzed the experimental distributions of
$i)$ the total range of the pion from the production point to
the stopping point, $ii)$ the pion range inside the detector,
$iii)$ the $\pi-\mu$ zenith scattering angle and $iv)$  the 
radial distance between the two particles in the bottom layer of scintillators.

The probability $A$ of events to be accepted, the tracking
reconstruction efficiency $E$ and the reconstruction efficiency $AE$ 
in cells of the parameter space 
${\cal P} (R_{\pi},\Delta\alpha,\Delta\varphi,D)$ have 
been evaluated using a Monte Carlo simulation of the detector.
This allows the calculation of the mean detection efficiency 
$\epsilon_r $ for events
in which the $\pi-\mu$ scattering angle $\Delta\alpha>45^o$.
We find $\epsilon_r = (21 \pm 2_{stat} \pm 4_{sys})\%$.
We have presented the detector-unfolded distributions of the $\pi-\mu$
zenith scattering angle $\Delta\alpha$, 
of the detector material thickness crossed by
the upgoing particles, and of the angular distribution of
charged particles emerging from the floor in the detector frame.
The measured rate of up-going pions with range $R_\pi > 25\ g\ cm^{-2}$
per down-going muon at the
Gran Sasso depth, corrected for the reconstruction efficiency of the 
MACRO apparatus, is $N_{\pi/\mu}= (10^{+4}_{-2.5})\ 10^{-5}$.

A comparison of the data with a Monte Carlo calculation
of up-going pion production in hard muon interactions shows good agreement 
in the measured pion production rate at the MACRO depth. 

The up-going particles generated
by interactions of down-going muons constitute a background source for
neutrino induced events.
For MACRO, the background rate is 
$(2.1 \pm 0.4_{stat} \pm 0.7_{sys})\quad events/year$
for the upward through-going sample with a muon energy 
threshold of $\sim 400\ MeV$, corresponding to  $200\ g\ cm^{-2}$ of detector. 
It is
$(8.8 \pm 0.9_{stat} \pm 0.8_{sys})\ events/year$
for the up-going stopping sample.
This background  is of the order of $1\%$
of the through-going muons and of the order of $10\%$ 
of the up-going stopping muons.

\section*{Acknowledgements}
We gratefully acknowledge the staff of the {\it Laboratori Nazionali
del Gran Sasso} and the invaluable assistance of the technical
staffs of all the participating Institutions.  For generous
financial contributions we thank the U.S.~Department of Energy,
the National Science Foundation, and the Italian {\it Istituto
Nazionale di Fisica Nucleare}, both for direct support and for FAI
grants awarded to non-Italian MACRO collaborators.

\end{document}